\documentclass[reprint,amsmath,amssymb,aps,floatfix]{revtex4-2}

\usepackage{graphicx}
\usepackage{dcolumn}
\usepackage{bm}
\usepackage{xcolor}
\usepackage{tikz}
\usetikzlibrary{quantikz}
\usepackage{subfigure}
\usepackage{braket}
\usepackage{url}

\newcommand{\qd}{q^{\dag}}
\newcommand{\cd}{c^{\dag}}
\newcommand{\ad}{a^{\dag}}

\begin{document}

\title{Parity Measurements using Dispersive Shifts for Surface Codes}

\author{Aneirin J. Baker}
\affiliation{SUPA, Institute of Photonics and Quantum Sciences,
Heriot-Watt University, Edinburgh EH14 4AS, United Kingdom}
\email{aneirin.baker@stfc.ac.uk}

\date{\today}

\begin{abstract}
Parity measurements are central to quantum error correction (QEC). In current implementations measurements of stabilizers are performed using a number of Controlled Not (CNOT) gates. This implementation suffers from an exponential decrease in fidelity as the number of CNOT gates increases thus the stabilizer measurements also suffer a severe decrease in fidelity and increase in gate time. Speeding up and improving the fidelity of this process will improve error rates of these stabilizer measurements thus increasing the coherence times of logical qubits. We propose a single shot method useful for stabilizer readout based on dispersive shifts. We show a possible set up for this method and simulate a 4 qubit system showing that this method is an improvement over the previous CNOT circuit in both fidelity and gate time. We find a fidelity of $99.8\%$ and gate time of $600$ ns using our method and investigate the effects of higher order Z interactions on the system. 
\end{abstract}

\maketitle

\section{Introduction}
As quantum computers progress, the need for quantum error correction (QEC) becomes ever clearer. Having a system whose lifetime and error rates exceed the levels of the individual physical qubits is essential for the future of quantum computing. Current developments within QEC have been focused on an implementation called the surface code \cite{krinner2021,Arute2019}, however other implementations have been explored such as the colour code and the Steane code \cite{Postler_2022,self2022,bluvstein2022}. At the heart of these implementations are measurements of stabilizers elements, these stabilizers are used to determine if an error has occurred in the system. Here we focus our attention on the surface code, a planar implementation of Kitaev’s Toric code \cite{Kitev2003}. This code is promising due to its high threshold for correctable errors in the system of around 1\% \cite{Raussendorf_2007}, and is amenable to implementation in 2D architectures, making it a promising contender to reach fault tolerant quantum computation.
Currently stabilizer measurements are measured via a series of CNOT gates that maps their eigenvalues onto the $Z$ eigenvalue of an ancilla qubit \cite{fowler2012}. In current superconducting implementations of surface codes the CNOT gates which make up these stabilizer measurements have fidelities of ~98.5\% \cite{krinner2021} when we consider that stabilizer measurements require between 2 and 4 CNOT gates simple calculations brings us to an estimated fidelity of 97\% and 94\% respectively for the stabilizer. Other implementations of the CNOT (or equivalent two qubit gates) in larger qubit devices are much lower than the required fidelity for effective stabilizer interactions \cite{krinner2021realizing,Sete2021,Arute2019,IBMq}.

\begin{figure}[ht]
\begin{quantikz}
  \lstick{$\ket{a}$} &  &\ctrl{1}  &\ctrl{1}   &\ctrl{1}   &\ctrl{1}       &\qw & \qw\\
\lstick{$\ket{q_1}$} &       &\targ{}   &\qw \vqw{1}&\qw \vqw{1}&\qw \vqw{1}    &\qw&\qw\\
 \lstick{$\ket{q_2}$}&       &\qw       &\targ{}    &\qw \vqw{1}&\qw\vqw{1}    &\qw&\qw\\
 \lstick{$\ket{q_3}$}&       &\qw       &\qw        &\targ{}    &\qw\vqw{1}     &\qw&\qw\\
\lstick{$\ket{q_4}$} &       &\qw       &\qw        &\qw        &\targ{}        &\qw&\qw\\
\end{quantikz}
\caption{Plaquette operator which is the basis for the error correcting codes. It comprises 4 CNOT gates with the ancilla as the target qubit.}
\label{fig:stabalizer_decomp}
\end{figure}
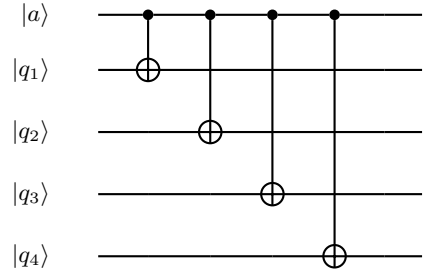
For QEC to be successful these error rates must be improved. A potential way, which has been investigated before in the context of the iToffoli gate \cite{baker2022,siddiqi2022}, is to implement a single shot measurement of this stabilizer. This single shot measurement has the advantage of not being effected by exponential scaling of fidelities for small stabilizer measurements and shows a modest improvement in gate time over its CNOT decomposition. We also find that it is easy to modify this single shot gate to other geometries of surface code.

We utilize dispersive shifts that can be engineered in superconducting circuits (SCCs) through the ZZ couplings between qubits \cite{Xu2021ZZGates,Collodo2019ObservationResonators,Sung2021RealizationCoupler,baker2022}. Here, we use these shifts to enact parity measurements on ensembles of qubits. We aim to find a regime where a subset of all two body dispersive shifts dominate and are of equal size. The subset we choose determines the operations we wish to perform.  For our example system we choose 4 qubit setup with 3 control qubits and one ancilla qubit. We choose to measure the odd parity of this system to show that the more complex situation of two drives is feasible.

To create a parity measurement we must detune the transitions $\ket{1001} \leftrightarrow \ket{1000},\ket{0101}\leftrightarrow \ket{0100} \text{ and }\ket{0011} \leftrightarrow \ket{0010}$ (where we have adopted the notation $\ket{\text{Qubit1, Qubit2, Qubit3, Ancilla}}$) from all other transitions of the ancilla. We then drive the ancilla at the detuned transition frequency ($\omega_{d} = \omega_a + \chi$, where $\chi$ is the dispersive shift we have engineered) only these three transitions will be driven and all the rest will be unaffected. If we also ensure that all other dispersive shifts are suppressed then we can use the same technique to drive higher order transitions for example, $\ket{1110} \leftrightarrow \ket{1111}$ with the drive $\omega_d = \omega_a + 3\chi$. Choosing the combination of drives correctly we can then generate a parity measurement on an ensemble of qubits. 

This paper is organised as follows in Section \ref{sec:PhyscialImp} we discuss the physical implementation of the model where we shall theoretically describe the system, in Section \ref{sec:simulation} we shall present the simulations of the of the model in the previous section. In Section \ref{sec:discussion} we shall discuss the implementation highlighting its advantages over other implementations and and discussing some of its drawbacks. 

\section{Physical implementation and model}\label{sec:PhyscialImp}
\begin{figure}[!ht]
    \centering
    \includegraphics[width=0.35\textwidth]{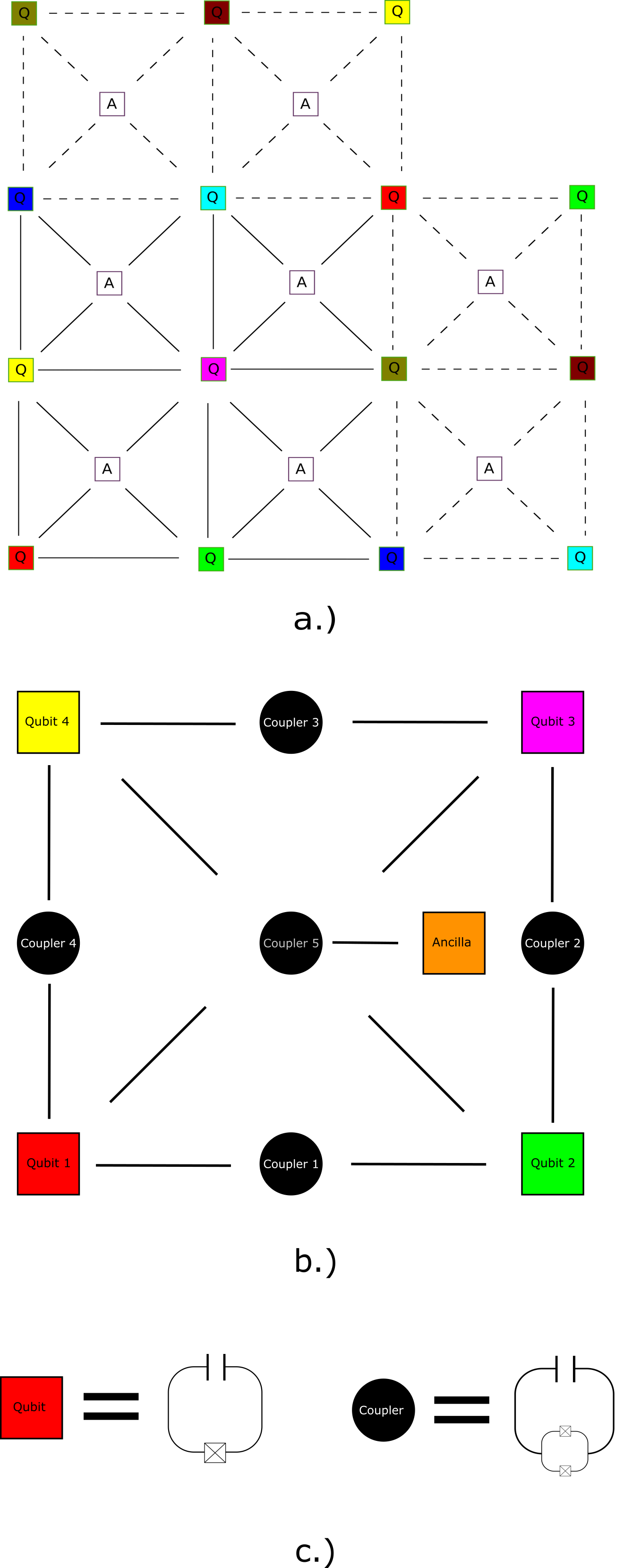}
    \caption{a.)Example of an effective lattice that tessellates the qubit frequencies over the surface to ensure that all qubits are sufficiently detuned from one another. We colour code the qubits to show which qubits would have different transition frequencies. 
    In this diagram the Q's represent the data Qubits and the A represents the ancilla qubits. b.) Representation of the circuit we shall be simulating in this chapter. Squares represent qubits and circles represent couplers. Each qubits is colour coded to represent a different transition frequency. This pattern can be tessellated over the plane so that all patches of qubits can have the correct dispersive shift to execute the parity check. c.) Legend for the circuit depicted above showing the qubits and couplers used in this chapter. In this chapter we use fixed frequency qubits and frequency qubit couplers. }
    \label{fig:lattice_and_diagram}
\end{figure}
\begin{figure*}[ht]
    \centering
    \includegraphics[width=0.9\textwidth]{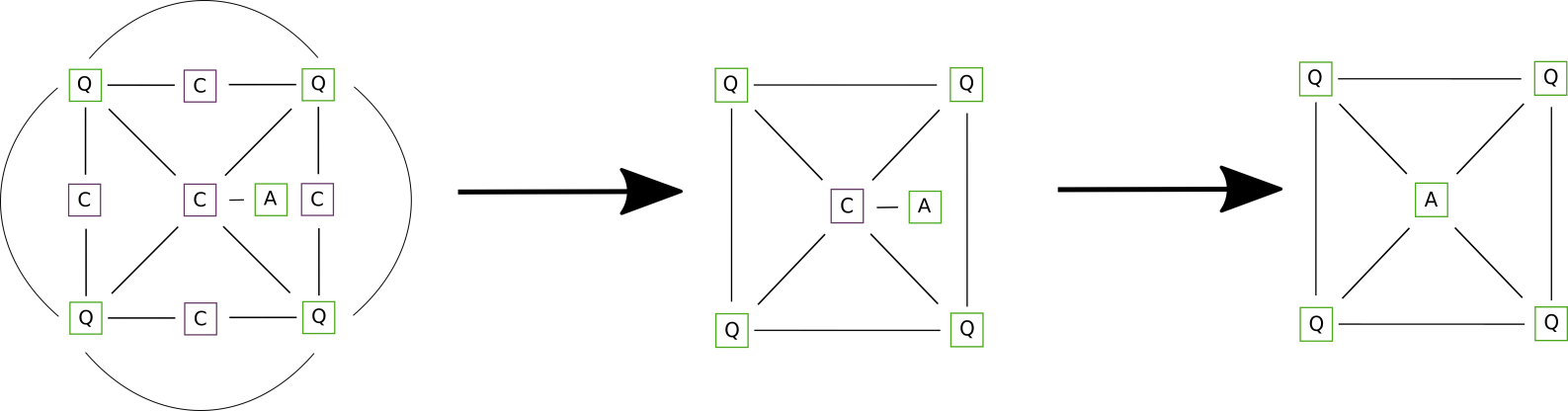}
    \caption{A pictorial illustration of the approximations we are making where green squares are the qubits, purple squares are the couplers (SQUIDS) and the black links are the capacitive couplings between the elements. At each stage the arrows represent the SW transformations outlined above. The aim of these approximations is to get an approximate theory of the system with the couplers eliminated and all the qubits connected to the ancilla.}
    \label{fig:SW_progression}
\end{figure*}

The readout of a stabilizer in QEC requires the data qubits to be connected to an ancilla - or measurement qubit. A recent realizations of simple QEC codes implement these gates using static capacitive coupling \cite{krinner2021}. This allowed for fast CZ gates to be implemented in 100ns. In our implementation we propose a lattice of qubits coupled via tunable couplers. This gives us the freedom to turn off interactions and adjust the strength of interaction by tuning the frequency of the couplers. This tunability also gives us access to different interaction regimes where we can perform gates specific to that regime. The lattice in Fig. \ref{fig:lattice_and_diagram} shows the qubit layout we propose, we note that this is very similar to current implementations \cite{Arute2019,Wu2021} thus we are expanding the capabilities of these types of Superconducting Circuits - SCCs.
We describe the Hamiltonian of a unit cell in terms of creation and annihilation operators and begin with the following model of the circuit in Fig. \ref{fig:SW_progression}
\begin{equation}
    H = \sum_{i=1}^4 H_0 (q_i) + \sum_{j=1}^5 H_0 (c_j) + H_0 (a) + H_{\text{int,q}} + H_{\text{int,c}}.
\end{equation}
Where we have defined
\begin{eqnarray}
    H_0 (X_i) &=& \omega_i X_i^{\dag} X_i + \frac{\alpha_i}{2} X_i^{\dag} X_i^{\dag} X_i X_i .
\end{eqnarray}
Here $X_i$($X_i^{\dagger}$) represents a general annihilation (creation) operator.
\begin{eqnarray}
    H_{\text{int,q}} &=& \sum_{i<j} g_{ij} (q_i - q_i^{\dag})(q_j - q_j^{\dag}),
    \\
    H_{\text{int,c}} &=& \sum_{i=1}^4 g_{i,ci} (q_i - q_i^{\dag})(c_i - c_i^{\dag}) 
    \\ \nonumber
    &+& g_{i+1,ci} (q_{i+1} - q_{i+1}^{\dag})(c_i - c_i^{\dag}) 
    \\ \nonumber
    &+& g_{i,c5} (q_i - q_i^{\dag})(c_5 - c_5^{\dag}) 
    \\ \nonumber
    &+& g_{a,c5} (a - a^{\dag})(c_5 - c_5^{\dag}) .
\end{eqnarray}
In the above Hamiltonian $q_i$ ($c_i$,$a$)/$q_i^{\dag}$ ($c_i^{\dag}$,$a^{\dag}$) represent the annihilation/creation operators for the qubits (couplers, ancilla) which obey the commutation relations
\begin{eqnarray}\label{eq:commutation_relations}
    &&[q_i,\qd_j] = [c_i,c^{\dagger}_j] = \delta_{ij} , \quad [a,a^{\dagger}] = 1.
\end{eqnarray}
All the other combinations of the operators are defined to be 0. In this model $\omega_i$/$\alpha_i$ ($\omega_{ci}$/$\alpha_{ci}$, $\omega_{a}$/$\alpha_{a}$) are the qubit (coupler, ancilla) transition frequencies and anharmonicities respectively. Here $g_{nm}$ denotes the coupling between qubit $n$ and $m$ (the exact form of the coupling can be found in Appendix \ref{sec:Appendix_Model_SW_transformation}) and $g_{i,cj}$ describes the coupling between the i-th qubit and the j-th coupler.
This model describes a full unit cell in the lattice in Fig. \ref{fig:lattice_and_diagram}. We can create an effective model for this by performing successive Schrieffer–Wolff transformations \cite{Bravyi2011Schrieffer-WolffSystems} eliminating sets of couplers along the way. At each point we ensure that the approximations made are small enough that we can make these approximations - see Appendix \ref{sec:Appendix_Model_SW_transformation} for further details. For the first approximation we eliminate the couplers along the edges of the unit cell with the transformation
\begin{equation}
    H \to \tilde{H} = e^{i S_{\text{edge}}} H e^{-i S_{\text{edge}}},
\end{equation}
with 
\begin{widetext}
\begin{equation}
S_{\text{edge}} = \sum_{i=1}^4 \bigg( \frac{g_{i,c_i}}{\Delta_{i,ci}} (\qd_i c_i - q_i \cd_i) - \frac{g_{i,c_i}}{\Sigma_{i,ci}} (\qd_i \cd_i - q_i c_i) 
+ \frac{g_{i+1,c_i}}{\Delta_{i+1,ci+1}} (\qd_{i+1} c_i - q_{i+1} \cd_i) 
- \frac{g_{i+1,c_i}}{\Sigma_{i+1,ci+1}} (\qd_{i+1} \cd_i - q_{i+1} c_i)\bigg).
\end{equation}
\end{widetext}
Here we have defined $\Delta_{i,j} = \omega_i - \omega_j$, $\Sigma_{i,j} = \omega_i + \omega_j$ and we have used a notation where the indices are defined modulo 4 to reflect the periodic nature of the system. We then apply a similar transformation
\begin{eqnarray}
    \tilde{H} \to \Bar{H} = e^{i S_{\text{center}}} \tilde{H} e^{-i S_{\text{center}}},
\end{eqnarray}
with the argument now 
\begin{eqnarray}
\nonumber
S_{\text{center}} &=& \sum_{i=1}^4 \frac{\tilde{g}_{i,c_5}}{\tilde{\Delta}_{i5}} (\qd_i c_5 - q_i \cd_5) - \frac{\tilde{g}_{i,c_5}}{\tilde{\Sigma}_{i5}} (\qd_i \cd_5 - q_i c_5) 
\\ 
&+& \frac{\tilde{g}_{a,c_5}}{\tilde{\Delta}_{a5}} (\ad c_5 - a \cd_5) - \frac{\tilde{g}_{a,c_5}}{\tilde{\Sigma}_{a5}} (\ad \cd_5 - a c_5),
\end{eqnarray}
to eliminate the central coupler. Here we have defined $\tilde{\Delta}_{i,j} = \tilde{\omega}_i - \tilde{\omega}_j$, $\tilde{\Sigma}_{i,j} = \tilde{\omega}_i + \tilde{\omega}_j$. This results in the final Hamiltonian of
\begin{eqnarray} \label{eq:final_hamiltonian}
    \Bar{H} &=& \sum_{i=1}^5 \Bar{H}_0 (c_i) + \sum_{j=1}^4 \Bar{H}_0 (q_i) 
    \\ \nonumber
    &+&\Bar{H}_{\text{int,q}} + O(\frac{g^2}{\Delta_{i,5}^2}),
    \\
    \Bar{H}_0 (X_i) &=& \Bar{\omega}_i X_i^{\dag} X_i + \frac{\Bar{\alpha}_i}{2} X_i^{\dag} X_i^{\dag} X_i X_i ,
    \\
    \Bar{H}_{\text{int,q}} &=& \sum_{i<j} \Bar{g}_{ij} (q_i - q_i^{\dag})(q_j - q_j^{\dag}).
\end{eqnarray}
Here $\Bar{\omega}_{n}$, $\Bar{\alpha}_{n}$ and $\Bar{g}_{nm}$ are shifted frequencies, nonlinearities and couplings - see appendix \ref{sec:Appendix_Model_SW_transformation} for the explicit expressions. 

We pause here to note that other implementations of this system are possible. It is conceivable that a multi qubit coupler \cite{Sameti2017SuperconductingCode,Frattini2017} could be used to create the same interactions as we have here and in a simpler manner. We choose this setup due to the flexibility gained from the tunable couplers and hence the ability to correct for the phase errors that could occur in the system.
\paragraph{Schrieffer–Wolff Commutation Errors}
As we have performed two successive SW transformations we must examine whether these two operators commute as these are still unitary transformations and so when we apply them to the Hamiltonian there will be some error accumulated if the two transformations do not commute. We see that these indeed do not commute since they both contain the qubit operators $q_i$ and $\qd_j$ which do not commute when $i=j$. According to the Baker–Campbell–Hausdorff formula the error due to these two operators will be proportional to $\frac{1}{2!}[S_{\text{edge}},S_{\text{Center}}]$ which will be proportional to $\frac{g^2}{2 \Delta^2}$. As we have been considering only Hamiltonians at less than second order in $g$ then we are safely able to ignore this error it has a strength of $\leq 1$ MHz.
\subsection{Dispersive Shifts and parity measurement}
With the model of the system now in place we derive the conditional shifts that form the basis of our gate. As done previously \cite{baker2022,Zhu2013CircuitRegime} we treat the interaction term as a perturbation to the system and apply perturbation theory with $V = \Bar{H}_{\text{int,q}}$ to determine the qubit transition frequencies up to $n$-th order in perturbation theory. Using these expressions we can determine the dispersive shifts present in the system. Specifically we are looking for the $n$-body dispersive shifts - defined in Appendix \ref{sec:Dispersive_Shifts}. In previous explorations of dispersive shifts only three body shifts were considered as were only, at most, three qubits coupled together. However, for our system and therefore extensions of our system we must consider $n$-body shifts. For example in the simulations below we need to consider 4-body dispersive shift in $n$-th order perturbation theory defined by
\begin{widetext}
\begin{align}
\nonumber
    \chi_{1234}^{\text{bare},\text{(n)}} &= (E_{\ket{1111}}^{\text{(n)}} - E_{\ket{0000}}^{\text{(n)}}) - (E_{\ket{1000}}^{\text{(n)}} - E_{\ket{0000}}^{\text{(n)}}) - (E_{\ket{0100}}^{\text{(n)}} - E_{\ket{0000}}^{\text{(n)}}) - (E_{\ket{0010}}^{\text{(n)}} - E_{\ket{0000}}^{\text{(n)}})
    - (E_{\ket{0001}}^{\text{(n)}} - E_{\ket{0000}}^{\text{(n)}})
    \\ 
    &= E_{\ket{1111}}^{\text{(n)}} - E_{\ket{1000}}^{\text{(n)}} - E_{\ket{0100}}^{\text{(n)}} - E_{\ket{0010}}^{\text{(n)}}- E_{\ket{0001}}^{\text{(n)}} + 3E_{\ket{0000}}^{\text{(n)}}.
\end{align}
\end{widetext}
We denote the above dispersive shift as the ``bare" shift since this shift also contains contributions from lower order dispersive shifts which can be expressed as
\begin{equation}\label{eqt:Dispersive_shifts_Full}
\chi_{1234}^{\text{(n)}} = \chi_{1234}^{\text{\text{bare},\text{(n)}}} - \sum_{i\neq j \neq k} \chi_{ijk}^{\text{\text{bare},\text{(n)}}} - \sum_{i \neq j} \chi_{ij}^{\text{(n)}}.
\end{equation}
Here $\chi_{ijk...}^{(m)}$ is the Dispersive shift on the qubits $ijk...$ to $m$-th order in perturbation theory and similarly $E_{\ket{n}}^{(m)}$ is the energy of state $\ket{n}$ to $m$-th order in perturbation theory. Using these expressions we can estimate the effects of these dispersive shifts to find a regime where we can suppress the unwanted interactions. 
\begin{widetext}
\begin{align}
    &\chi_{1234}^{(2)} =
    \\ \nonumber
    & \frac{-4 g_{1,a}^2}{\alpha_1+\alpha_a+\Sigma_{1,a}}
    +\frac{2 g_{1,a}^2}{\alpha_1+\Sigma_{1,a}}
    -\frac{2 g_{1,a}^2}{\alpha_1+\Delta_{1,a}}
    +\frac{2 g_{1,a}^2}{\alpha_a+\Sigma_{1,a}}
    -\frac{2 g_{1,a}^2}{\alpha_a -\Delta_{1,a}}
    -\frac{g_{1,a}^2}{\Sigma_{1,a}}
    -\frac{g_{1,a}^2}{\Delta_{1,a}}
    +\frac{g_{1,a}^2}{\Delta_{1,a}}
    +\frac{g_{1,a}^2}{\Sigma_{1,a}}
    \\ \nonumber
    &-\frac{4 g_{2,a}^2}{\alpha_2+\alpha_a+\Sigma_{2,a}}
    +\frac{2 g_{2,a}^2}{\alpha_2+\Sigma_{2,a}}
    -\frac{2 g_{2,a}^2}{\alpha_2+\Delta_{2,a}}
    +\frac{2 g_{2,a}^2}{\alpha_a+\Sigma_{2,a}}
    -\frac{2 g_{2,a}^2}{\alpha_a- \Delta_{2,a}}
    -\frac{g_{2,a}^2}{\Sigma_{2,a}}
    -\frac{g_{2,a}^2}{\Delta_{2,a}}
    +\frac{g_{2,a}^2}{\Delta_{2,a}}
    +\frac{g_{2,a}^2}{\Sigma_{2,a}}
    \\ \nonumber
    &-\frac{4 g_{3,a}^2}{\alpha_3+\alpha_a+\Sigma_{3,a}}
    +\frac{2 g_{3,a}^2}{\alpha_3+\Sigma_{3,a}}
    -\frac{2 g_{3,a}^2}{\alpha_3+\Delta_{3,a}}
    +\frac{2 g_{3,a}^2}{\alpha_a+\Sigma_{3,a}}
    -\frac{2 g_{3,a}^2}{\alpha_a-\Delta_{3,a}}
    -\frac{g_{3,a}^2}{\Sigma_{3,a}}
    -\frac{g_{3,a}^2}{\Delta_{3,a}}
    +\frac{g_{3,a}^2}{\Delta_{3,a}}
    +\frac{g_{3,a}^2}{\Sigma_{3,a}}
    \\ \nonumber 
    &+ \text{Counter Rotating Terms},
    \\ \nonumber
    &= \chi_{12}^{(2)} + \chi_{13}^{(2)} + \chi_{14}^{(2)} + \text{Counter Rotating Terms}.
\end{align}
\end{widetext}
To second order we find that these higher order dispersive shifts can be decomposed into second order terms with cross terms which are proportional to the counter rotating terms. 
These counter rotating terms will not contribute much to the dispersive shifts since they are proportional to $\propto \frac{1}{\omega_i + \omega_j}$ which, for our parameters, will be small. These are the higher order shifts present in the summation terms of Eq. (\ref{eqt:Dispersive_shifts_Full}).

Using these expressions we find parameter regimes (by sweeping over a large parameter set within the dispersive regime) where all the pairwise dispersive shifts that include the ancilla are equal, any higher order interaction are highly suppressed and finally all other two body shifts are highly suppressed - see section \ref{sec:Appendix_DispersiveShifts} for exact shifts. Once we have found these regimes we can simulate the dynamics of the system choosing the number of drives and their frequencies according to the parity measurement we wish to execute.

\subsection{Simulations}\label{sec:simulation}
We now simulate a system with 3 data qubits,one ancilla qubit and two drives to demonstrate the proposed mechanism. As the Hamiltonian we derived has different behaviours for different initial states we apply two drives to the system to measure the parity. These drives will excite the central ancilla qubits if and only if the data qubits are in the parity state we wish to measure. To select a specific parity state we choose the frequencies of the external drives to match the shifts we have induced in the system. To measure the effectiveness of these simulations we define the ideal evolution to be the evolution  where the transitions $\ket{0000} \leftrightarrow \ket{1000}$, $\ket{0110} \leftrightarrow \ket{1110}$, $\ket{0101} \leftrightarrow \ket{1101}$ and $\ket{0011} \leftrightarrow \ket{1011}$ occur, and all other states are unchanged. Using the definition of process fidelity as $F_p(U_1,U_2) = |\text{Tr} (U_1^\dag , U_2)|/d$ ($d=16$ being the dimension of the Hilbert space) we can quantify how well our scheme realizes a stabilizer measurement and so compare it to other implementations of parity/stabilizer measurements. We choose the parameters shown in Table. \ref{tab:unitary_parameters} such that all the higher order shifts are suppressed and only the shifts we require are active. These parameters result in dispersive shifts of $\chi = \chi_{12} = \chi_{13} = \chi_{14} = -5$ MHz while all other dispersive shifts have absolute values $\leq 0.4$ MHz.
\begin{figure}[ht]
    \centering
    \includegraphics[width=0.4\textwidth]{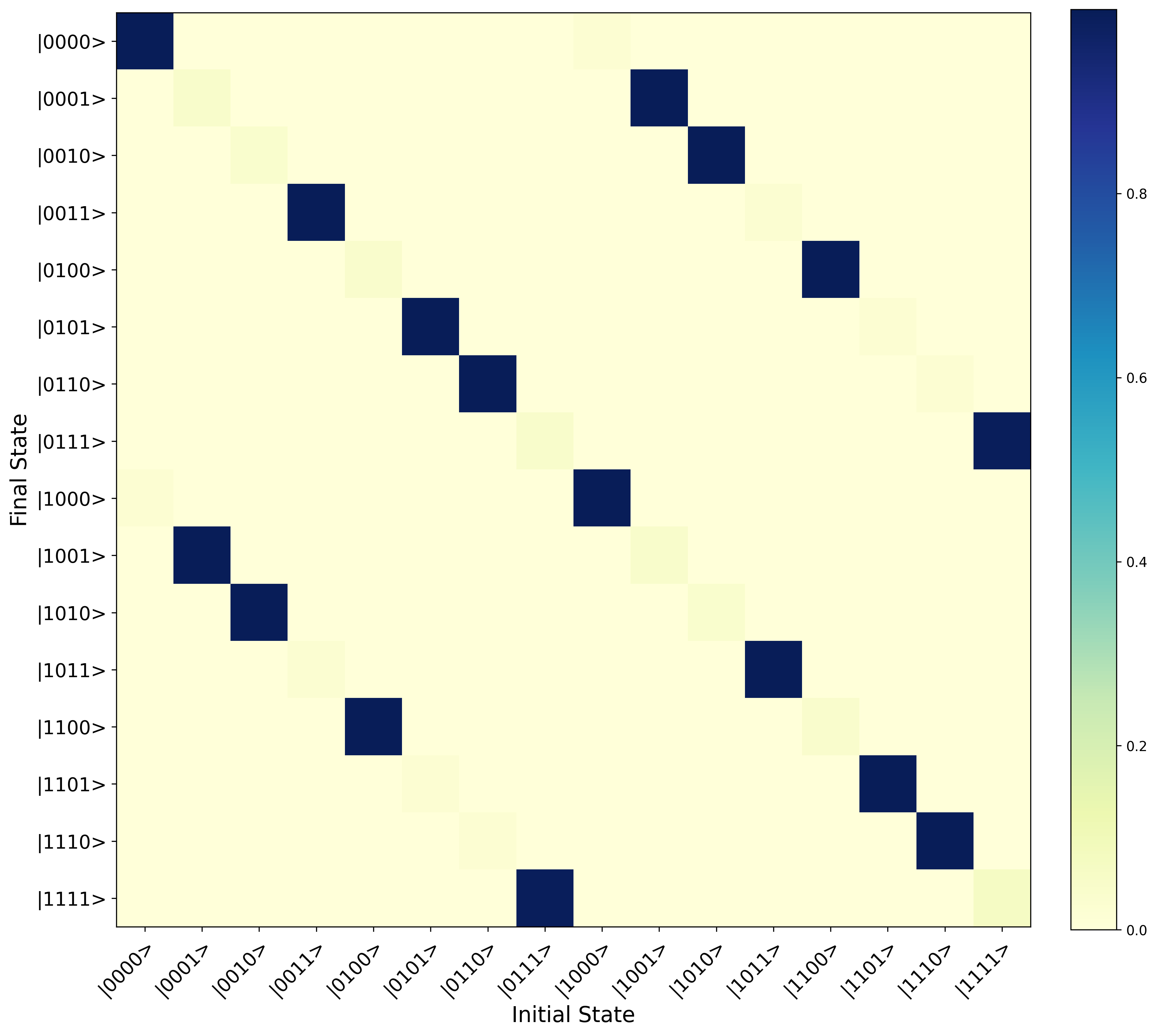}
    \caption{We simulate the system in Eq. (\ref{eq:final_hamiltonian}) using 3 data qubits and 1 measurement qubit. Two drives are applied to the system with frequencies of $\omega_{d1} = \Bar{\omega}_a + \chi$ and $\omega_{d2} = \Bar{\omega}_a + 3\chi$, these drives are what executes the parity measurement. There is still some small amount of population left in the odd parity states. This is due to the non-zero unwanted dispersive shifts. Above we show the absolute values of the unitary produced from this simulation.}
    \label{fig:Two_drive_unitary}
\end{figure}

\begin{table}[ht]
    \begin{center}
        \begin{tabular}{||c c||} 
         \hline
         Parameter & Value $[GHz]$ \\ [0.5ex] 
         \hline\hline
         $[\omega_1 , \omega_2 , \omega_3 , \omega_4]$
         & $[4.95,5.28,5.4,5.48]$  \\ 
         \hline
         $[\alpha_1 , \alpha_2 , \alpha_3 , \alpha_4]$ & $[-0.3,-0.2,-0.2,-0.19]$  \\
         \hline
         $[g_{12} , g_{13} , g_{14} ]$ & $[0.02165,0.032,0.0385]$ \\
         \hline
         $[g_{23} , g_{24} , g_{34} ]$ & $[0.001,0.001,0.001]$ \\
         \hline
         $[\Omega_1 , \Omega_2 ]$ & $[0.00159,0.00159]$ \\
         \hline
         $[\omega_{d1} , \omega_{d2} ]$ & $[4.938,4.929]$ \\
         \hline
        \end{tabular}

    \caption{Table outlining the parameters used in the simulations in Fig \ref{fig:Two_drive_unitary}. In this system $\omega_1$ represents the ancilla/measurement qubit and qubits 2-4 represent data qubits. This system produces a dispersive shift of $\chi =\chi_{12} = \chi_{13} = \chi_{14} = -5 MHz$. Whilst this is small it is enough to detune the transitions frequencies such that no other excitations occur when we apply the drives $\omega_{d1} = \Bar{\omega}_a + \chi$ and $\omega_{d2} = \Bar{\omega}_a + 3\chi$.}
     \label{tab:unitary_parameters}
    \end{center}
\end{table}

Fig. \ref{fig:Two_drive_unitary} shows the resulting unitary from simulations of  Eq. (\ref{eq:final_hamiltonian}) with two drives where $\omega_{d1} = E_{\ket{1100}} - E_{\ket{0100}}$ and $\omega_{d2} = E_{\ket{1111}} - E_{\ket{0111}}$. The overall process fidelity of this operation was calculated to be $F_p = 99.8\%$ with a total execution time of $t_{\text{gate}} = 600$ ns. 

\paragraph{Errors:}
The reduction in fidelity of $\approx 0.2 \%$ can be attributed to the population leakage to other states. We estimate a $0.1\%$ fidelity reduction due to leakage to states within the computational subspace and the other $0.1\%$ due to leakage to higher order states.
\\
As the gate time of our operation is reasonably large we estimate that decoherence may play a significant role in reducing fidelity. Current dephasing and detuning times are approaching the $100 \mu s$ mark \cite{IBMq,gold2021}. Whilst these coherence times are constantly improving, these current values would reduce the fidelity of our operation to $\approx 99.2\%$ (using a $100 \mu$s $T_1$ time), which is still much higher than the fidelities currently possible through the CNOT decomposition method.

In our simulations and fidelity calculations we have assumed perfect correction of the phases accumulated by the dispersive shifts we are using. In reality this is not the case, these gates have finite fidelities. However a modest change to our parameters solves this problem rather elegantly. It is feasible that we can arrange the parameters such that $\chi t_{\text{gate}} = 2 \pi$ (the disperisve shift between the ancilla and the data qubits). For our case this would involve a modest increase in the dispersive shift to $-7.5$ MHz and an increase in the gate time to $~830$ns however this would completely negate the issue of accumulated phase for this system. 

\section{Discussion}\label{sec:discussion}
We have shown that it is possible to execute higher order parity measurements in a single step. This technique can be extended further by adding extra qubits to the unit cell, connected to their neighbours and ancilla via tunable couplers. Our single shot measurement strategy eliminates the need for multiple CNOT gates reducing the effect of multiplying gate errors, this reduction in gate errors comes with the potential for faster parity gates. We also reduce other errors such as errors occurred during the idle time whilst other qubits wait for the CNOT gates to execute. 

This technique can be extended to more qubits in two different ways. Firstly  we can ensure that all two body shifts that are required for the larger parity measurements are dominant and equal whilst also ensuring that other (unwanted) shifts are sufficiently suppressed. We then pick the correct number of drives for the parity measurement required. In doing this we would have to suppress all Z-shifts at orders between 2 and n (where n is the number of qubits in the cluster we are measuring) provided the system is in the dispersive regime. 

Alternatively we can concatenate these lower order parity measurements together to create a higher order parity measurement. This follows the same mechanism as the current parity decomposition to CNOT gates where the CNOT gates measure the individual parity of each qubit-ancilla pair. Hence, if we know the parity of two ``sub clusters" we can infer the parity of the entire cluster, making this proposal useful for much larger parity measurements.

For the operation to retain the high fidelity we have predicted, we require a balance between the size of the drive and the gate time. This balance is determined by the coherence time of the qubits. We require that the gate be short enough that decoherence doesn't play a significant role in the reduction of fidelity but long enough that the drive can remain small enough that we remain in the weak driving regime. We also require that the parameters be chosen such that higher order interactions are suppressed so that the transitions we aim to execute are not significantly detuned from the drives. 

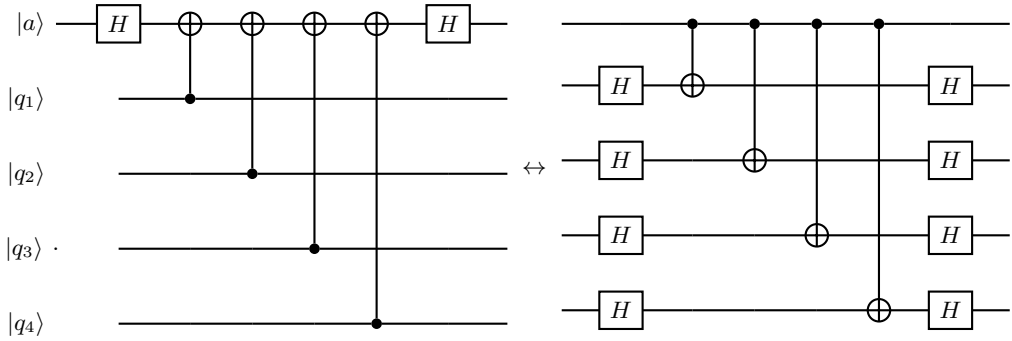
\begin{figure*}
\begin{equation}
\nonumber
\begin{quantikz}
\lstick{$\ket{a}$}& \gate{H}  & \targ{}   & \targ{}   & \targ{}   & \targ{}   & \gate{H}  & \qw \\
\lstick{$\ket{q_1}$}& \ghost{H} & \ctrl{-1} & \qw       & \qw       & \qw       & \qw       & \qw\\
\lstick{$\ket{q_2}$}& \ghost{H} & \qw       & \ctrl{-2} & \qw       & \qw       & \qw       & \qw\\.
\lstick{$\ket{q_3}$}& \ghost{H} & \qw       & \qw       & \ctrl{-3} & \qw       & \qw       & \qw\\
\lstick{$\ket{q_4}$}& \ghost{H} & \qw       & \qw       & \qw       & \ctrl{-4} & \qw       & \qw
\end{quantikz} 
\leftrightarrow
\begin{quantikz}
& \qw       & \ctrl{1}  & \ctrl{2}  & \ctrl{3}  & \ctrl{4}  & \qw       & \qw \\
& \gate{H}  & \targ{}   & \qw       & \qw       & \qw       & \gate{H}  & \qw \\
& \gate{H}  & \qw       & \targ{}   & \qw       & \qw       & \gate{H}  & \qw\\
& \gate{H}  & \qw       & \qw       & \targ{}   & \qw       & \gate{H}  & \qw\\
& \gate{H}  & \qw       & \qw       & \qw       & \targ{}   & \gate{H}  & \qw
\end{quantikz} 
\label{fig:X_stabalizer_decomp}
\end{equation}
\caption{Circuit diagrams showing the transformation of an X stabilizer (RHS) into a Z stabilizer (LHS) and vice versa. The transformation inverts the target of a CNOT gate through the use of Hadamard interactions. On the left-hand side the effect of the middle four CNOT gates is to produce parity measurements, the same measurement we have been developing here. This circuit represents the Z stabilizer measurement circuit, on the right we have the X stabilizer measurement. Shown here is the ability to transform between the two. This transformation allows us to create both of the possible stabilizers and others in between. One can imagine only performing the CNOT transformation applied here to two of the qubits thus creating a $XZZX$ stabilizer.}
\end{figure*}

Whilst we have only discussed the Z stabilizer measurements it is possible to turn these into an X stabilizer measurement. We can use the fact that one can exchange the roles of control and target qubits in CNOT gates by sandwiching it between Hadamard gates. Using this relation we can then turn the X syndrome measurement into a parity measurement with Hadamard gates applied to the data qubits instead of the measurement qubits. This results in the circuit in Eq.~(\ref{fig:X_stabalizer_decomp}). With this transformation it is easy to see that the method outlined above can be used to perform both X and Z stabilizer measurements.

\begin{figure*}
\centering
\begin{quantikz}
\lstick{$\ket{q_1}$}& \qw & \gate[3]{\text{3-Parity}} & \qw & \qw & \qw & \qw \\
\lstick{$\ket{q_2}$}& \qw & & \qw& \qw & \qw & \qw \\
\lstick{$\ket{q_3}$}& \qw & & \qw& \qw & \qw & \qw \\
\lstick{$\ket{q_4}$}& \qw & \qw & \qw & \qw & \gate[2]{\text{2-Parity}} & \qw\\
\lstick{$\ket{q_5}$}& \qw & \qw & \qw & \qw & & \qw \\
\lstick{$\ket{a}$}& \qw & \ctrl{-3} & \qw & \qw & \ctrl{-1} & \meter{}
\end{quantikz}
\caption{Example of the concatenation of multiple parity gates together to create a five fold parity measurement gate. We have broken the gate down into two gates a three qubit parity gates (3-parity gate) and a single two qubit parity gate (2-parity gate). This is a natural extension of the CNOT method outlined in Figure \ref{fig:stabalizer_decomp} where we can consider the CNOT gate as a single parity measurement gate or 1-parity measurement gate.}
\end{figure*}
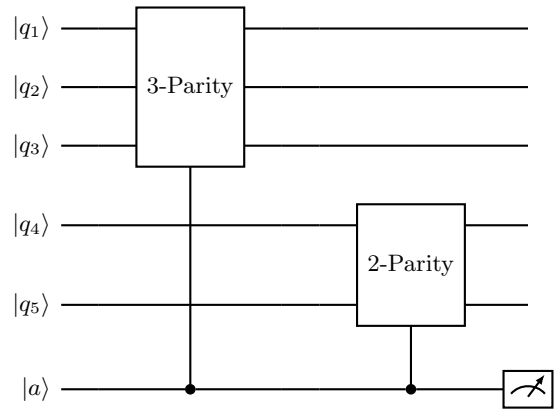
\section{Conclusion}
In conclusion, we have analysed the effectiveness of a single shot stabilizer measurement within superconducting circuits. We have shown the fidelity to be well above current estimates for CNOT decomposition's that are suggested in most implementations of quantum error correction. We discussed the errors associated with this system highlighting the effects of higher order Z interactions which could be corrected for in this system due to the connectivity of the circuit we have proposed. This system uses tunable couplers to couple the qubits together giving the system a large amount of tunability that can be used to execute different types of stabilizer interactions along with the ones we have proposed. Whilst this system shows the use of higher order Z interactions for quantum error correction the explanation of this through perturbation theory may be lacking slightly as 4th order perturbation theory has recently been shown to fail for certain values of the coupler frequency. A more thorough analysis of the higher order interactions is needed to fully understand these interactions and their uses.

After the conclusion of this work the authors were made aware of an independent work \cite{Reagor2022} based on a very similar effect.

\begin{acknowledgments}
This project has received funding from EPSRC DTP grant EP/R513040/1, I would also like to thank my supervisor M.Hartmann for his advise on this work.
\end{acknowledgments}

\clearpage
\newpage
\appendix

\section{Model}
\subsection{SW transformation}\label{sec:Appendix_Model_SW_transformation}
When performing the SW approximations we eliminate terms which we deem too small to contribute to the overall dynamics. Here we show the full transformations highlighting the approximations and calculating the contributions that these will make and showing that they will be small enough not to contribute. We keep terms only to second order in the coupling as higher terms are considered small enough to not contribute to the dynamics of the system.
\subsection{Elimination of edge couplers}
We begin by eliminating the coupler connecting the qubits along the edges of the unit cell. We use the transformation stated in the main text
\begin{eqnarray}\label{eq:SW_expansion_commutator}
    H \to \tilde{H} &=& e^{i S_{\text{edge}}} H e^{-i S_{\text{edge}}}
    \\ \nonumber
    e^{i S_{\text{edge}}} &H& e^{-i S_{\text{edge}}} = H + [S_{\text{edge}},H] 
    \\ \nonumber
    &+& \frac{1}{2!} [S_{\text{edge}},[S_{\text{edge}},H]] + ...
\end{eqnarray}
with 
\begin{widetext}
\begin{align}
\\ \nonumber
S_{\text{edge}} = \sum_{i=1}^4 \bigg( \frac{g_{i,c_i}}{\Delta_{i,ci}} (\qd_i c_i - q_i \cd_i) - \frac{g_{i,c_i}}{\Sigma_{i,ci}} (\qd_i \cd_i - q_i c_i) 
+ \frac{g_{i+1,c_i}}{\Delta_{i+1,ci+1}} (\qd_{i+1} c_i - q_{i+1} \cd_i) 
- \frac{g_{i+1,c_i}}{\Sigma_{i+1,ci+1}} (\qd_{i+1} \cd_i - q_{i+1} c_i)\bigg).
\end{align}
\end{widetext}
Here we have defined $\Delta_{i,j} = \omega_i - \omega_j$, $\Sigma_{i,j} = \omega_i + \omega_j$. We also have defined $(q_i, c_i,a) / (\qd_i, \cd_i,a^{\dagger})$ to be the qubit, coupler and ancilla annihilation/creation operators. In addition, we have defined $g_{nm}$ to denote the coupling between qubit $n$ and $m$ (the exact form can be found in the supplementary material) and $g_{i,cj}$ describes the coupling between the i-th qubit and the j-th coupler. For later use we shall also define the transition frequencies and anharmonicities for the ith qubit, coupler and the ancilla to be $\omega_i, \omega_{ci}$ and $\omega_a$.

The full calculation is long and obtuse, so we will only state the final result in this section. The calculation is similar to many other SW transformations to eliminate tunable couplers, which can e.g. be found in \cite{Sameti2017SuperconductingCode,Yan2018}. Keeping terms up to second order in the coupling we obtain the Hamiltonian
\begin{eqnarray}
    \tilde{H} &=& \sum_{i=1}^4 \tilde{H}_0 (q_i) + \sum_{j=1}^5 \tilde{H}_0 (c_j) + \tilde{H}_0 (a) 
    \\ \nonumber
    &+& \tilde{H}_{\text{int,q}} + \tilde{H}_{\text{int,c}},
\end{eqnarray}
with the functions now defined as 
\begin{eqnarray}
    \tilde{H}_0 (X_i) &=& \tilde{\omega}_i X_i^{\dag} X_i + \frac{\tilde{\alpha}_i}{2} X_i^{\dag} X_i^{\dag} X_i X_i,
    \\
    \tilde{H}_{\text{int,q}} &=& \sum_{i<j} \tilde{g}_{ij} (q_i - q_i^{\dag})(q_j - q_j^{\dag}),
    \\
    \tilde{H}_{\text{int,c}} &=& \tilde{g}_{i,c5} (q_i - q_i^{\dag})(c_5 - c_5^{\dag}) ,
    \\ \nonumber
    &+& \tilde{g}_{ci,c5} (c_i - \cd_i)(c_5 \cd_5)
    \\ \nonumber
    &+& \tilde{g}_{ci+1,c5} (c_{i+1} - \cd_{i+1})(c_5 - \cd_5).
\end{eqnarray}
Where
\begin{eqnarray} \label{eqt:Dressed_variables_tilde}
\tilde{\omega}_{i} &=& \omega_{i} +  g_{i,ci}^2 (\frac{1}{\Delta_{i,ci}} +\frac{1}{\Sigma_{i,ci}})
\\ \nonumber
&+& g_{i-1,ci-1}^2 (\frac{1}{\Delta_{i-1,ci-1}} + \frac{1}{\Sigma_{i-1,ci-1}}),
\\ \nonumber
\tilde{\omega}_{ci} &=& \omega_{ci} +  g_{i,ci}^2 (\frac{1}{\Delta_{i,ci}} + \frac{1}{\Sigma_{i,ci}})
\\ \nonumber
 &+& g_{i+1,ci+1}^2 (\frac{1}{\Delta_{i+1,ci+1}} + \frac{1}{\Sigma_{i+1,ci+1}}),
\\ \nonumber
\tilde{\omega}_a &=& \omega_a ,
\\ \nonumber
\tilde{g}_{ij} &=& g_{ij} +  g_{i,ci} g_{j,cj}\left(\frac{1}{\Delta_{i,ci}} + \frac{1}{\Delta_{j,cj}} - \frac{1}{\Sigma_{i,ci}} - \frac{1}{\Sigma_{j,cj}} \right), 
\\ \nonumber
\tilde{g}_{ci,c5} &=& g_{i,ci} g_{i,c5} \left(\frac{1}{\Delta_{i,ci}} + \frac{1}{\Sigma_{i,ci}} \right),
\\ \nonumber
\Tilde{\alpha}_i &\approx& \alpha_i,
\quad \text{,} \quad
\Tilde{\alpha}_{ci} \approx \alpha_{ci},
\quad \text{,} \quad
\Tilde{\alpha}_{a} \approx \alpha_{a}.
\end{eqnarray}
Here we have defined the shifted or ``first-dressed" system variables $\tilde{\omega},\tilde{\alpha}$ and $\tilde{g}$ where these represent the approximate transition frequency, anharmonicity and coupling of the new dressed system.
\\
We note that there is some effective coupling between the couplers and the central coupler but this term in proportional to $g_{i,ci} g_{i+1,c5}(\frac{1}{\Delta_{i,ci}} + \frac{1}{\Sigma_{i,ci}})$. Estimating this strength we find that for our parameters this coupling will have a reasonably large interaction strength. However the couplers should never gain any excitations as they are detuned from the qubits and are far detuned from any drive.
\\
We also find counter rotating terms $c_i c_i + c_i^{\dagger} c_i^{\dagger}$ that have strength given by $\frac{g_{i,ci}^2}{\Delta_{i,ci}}$. In a rotating frame these operators will rotate at the sum of their transition frequencies thus their dynamics will not affect the system and will average out to no contribution in the course of this interaction. 
\subsection{Elimination of central coupler}
Now that we have eliminated the edge couplers we move to eliminate the central coupler. This is again done with a Schrieffer–Wolff transformation but this time with the argument 
\begin{widetext}
\begin{align}
S_{\text{center}} = \sum_{i=1}^4 \frac{\tilde{g}_{i,c_5}}{\tilde{\Delta}_{i5}} (\qd_i c_5 - q_i \cd_5) - \frac{\tilde{g}_{i,c_5}}{\tilde{\Sigma}_{i5}} (\qd_i \cd_5 - q_i c_5) 
+ \frac{\tilde{g}_{a,c_5}}{\tilde{\Delta}_{a5}} (\ad c_5 - a \cd_5) - \frac{\tilde{g}_{a,c_5}}{\tilde{\Sigma}_{a5}} (\ad \cd_5 - a c_5).
\end{align}
\end{widetext}
We perform the transformation 
\begin{eqnarray}
    \tilde{H} \to \Bar{H} = e^{i S_{\text{center}}} \tilde{H} e^{-i S_{\text{center}}},
\end{eqnarray}
using the commutations relations in Eq.(\ref{eq:commutation_relations}) as stated in the main text and the commutator expansion Eq. \ref{eq:SW_expansion_commutator} we obtain
\begin{equation}
[S_{\text{center}},\tilde{H}]
\end{equation}
\begin{align}
= &- \sum_{i=1}^4 \tilde{g}_{i,c5} (q_i \cd_5 + \qd_i c_5) + \tilde{g}_{i,c5} (q_i c_5 + \qd_i \cd_5)
\\ \nonumber
&- \tilde{g}_{a} (a \cd_5 + \ad c_5) + \tilde{g}_{a} (a c_5 + \ad \cd_5)
\\ \nonumber
&-\sum_{i=1}^4 \sum_{j<k} (\frac{\tilde{g}_{i,c5} \tilde{g}_{jk}}{\tilde{\Delta}_{i,c5}} + \frac{\tilde{g}_{i,c5} \tilde{g}_{jk}}{\tilde{\Sigma}_{i,c5}})((q_k - \qd_k)(c_5 - \cd_5) \delta_{ij} 
\\\nonumber
&+ (q_j - \qd_j)(c_5 - \cd_5) \delta_{ik})
\\\nonumber
&+\sum_{i,j} \tilde{g}_{i,c5} \tilde{g}_{j,c5}(\frac{1}{\tilde{\Delta}_{i,c5}} - \frac{1}{\tilde{\Sigma}_{i,c5}})(q_j - \qd_j)(q_i - \qd_i)
\\\nonumber
&- (c_5 - \cd_5)^2
\\\nonumber
&+\sum_{i} \tilde{g}_{i,c5} \tilde{g}_{a}(\frac{1}{\tilde{\Delta}_{i,c5}} - \frac{1}{\tilde{\Sigma}_{i,c5}})(a-\ad)(q_i - \qd_i)
\\\nonumber
&+ \tilde{g}_a^2(\frac{1}{\tilde{\Delta}_{a,c5}} - \frac{1}{\tilde{\Sigma}_{a,c5}})((a-\ad)^2-(c_5 - \cd_5)^2).
\end{align}
\begin{equation}
[S_{\text{center}},[S_{\text{center}},\tilde{H}]]
\end{equation}
\begin{align}
\nonumber
&-\sum_{i,j} \tilde{g}_{i,c5} \tilde{g}_{j,c5} (q_i + \qd_i)(q_j+ \qd_j) (\frac{1}{\Sigma_{i,c5}} + \frac{1}{\Delta_{i,c5}}) 
\\ \nonumber
&-\sum_{i,j} \tilde{g}_{i,c5} \tilde{g}_{j,c5} (\frac{1}{\Sigma_{i,c5}} + \frac{1}{\Delta_{i,c5}})\delta_{ij} ((c_5 \cd_5 - \cd_5 c_5) + (c_5 c_5 + \cd_5 \cd_5))
\\ \nonumber
&-\sum_{j} \tilde{g}_{a} \tilde{g}_{j,c5} (\frac{1}{\Sigma_{a,c5}} + \frac{1}{\Delta_{a,c5}})(a-\ad)(a-\ad) 
\\ \nonumber
&-\sum_{j} \tilde{g}_{a} \tilde{g}_{j,c5} (\frac{1}{\Sigma_{a,c5}} + \frac{1}{\Delta_{a,c5}}) (c_5 + \cd_5)^2 
\\ \nonumber
&-\sum_{i,j}\tilde{g}_{a,c5} \tilde{g}_{i,c5} (\frac{1}{\Sigma_{a,c5}} + \frac{1}{\Delta_{a,c5}}) ((q_i \ad + \qd_i) - (q_i a - \qd_i \ad))
\\ \nonumber
&-\sum_{i,j}\tilde{g}_{a,c5} \tilde{g}_{j,c5} (\frac{1}{\Sigma_{j,c5}} + \frac{1}{\Delta_{j,c5}})((\qd_j a + \ad q_j) + (a q_j - \ad \qd_j)).
\\ \nonumber
\end{align}
with these calculations we find the final Hamiltonian to be
\begin{align} \label{eq:final_hamiltonian_app}
    &\Bar{H} = \sum_{i=1}^5 \Bar{H}_0 (c_i) + \sum_{j=1}^4 \Bar{H}_0 (q_i) +\Bar{H}_0 (a)
    \\ \nonumber
    &+\Bar{H}_{\text{int,q}} + O(\frac{g^2}{\Delta^2}),
    \\\nonumber
    &\Bar{H}_0 (X_i) = \Bar{\omega}_i X_i^{\dag} X_i + \frac{\Bar{\alpha}_i}{2} X_i^{\dag} X_i^{\dag} X_i X_i,
    \\\nonumber
    &\Bar{H}_{\text{int,q}} = \sum_{i<j} \Bar{g}_{ij} (q_i - q_i^{\dag})(q_j - q_j^{\dag}).
\end{align}
Where we define
\begin{eqnarray} \label{eqt:Dressed_variables_bar}
\Bar{\omega}_{qi} &=& \tilde{\omega}_{qi} +  \tilde{g}_{i,c5}^2 (\frac{1}{\tilde{\Delta}_{i,c5}} + \frac{1}{\tilde{\Sigma}_{i,c5}}), 
\\ \nonumber
\Bar{\omega}_{c5} &=& \tilde{\omega}_{c5} + \sum_{i=1}^4 \tilde{g}_{i,c5}^2 (\frac{1}{\tilde{\Delta}_{i,c5}} + \frac{1}{\tilde{\Sigma}_{i,c5}}), 
\\ \nonumber
\omega_a &=& \tilde{\omega}_a+ g_{a,c5}\left(\frac{1}{\tilde{\Delta}_{a,c5}} + \frac{1}{\tilde{\Sigma}_{a,c5}}\right),
\\ \nonumber
\Bar{g}_{ij} &=& \tilde{g}_{ij} +  \tilde{g}_{i,c5} \tilde{g}_{j,c5}\left(\frac{1}{\tilde{\Delta}_{i,c5}} + \frac{1}{\tilde{\Delta}_{j,c5}} - \frac{1}{\tilde{\Sigma}_{i,c5}} - \frac{1}{\tilde{\Sigma}_{j,c5}} \right), 
\\ \nonumber 
\Bar{\alpha}_i &\approx& \tilde{\alpha}_i
\quad \text{,} \quad \Bar{\alpha}_a \approx \tilde{\alpha}_a .
\end{eqnarray}
Here we have $\tilde{\Delta}_{i,j} = \tilde{\omega}_i - \tilde{\omega}_j$, $\tilde{\Sigma}_{i,j} = \tilde{\omega}_i + \tilde{\omega}_j$. In addition, we now have the ``second-dressed" variables being $\Bar{\omega}_{n}$ and $\Bar{\alpha}_{n}$ and $\Bar{g}_{nm}$ are shifted frequencies, anharmonicities and couplings respectively.
\section{Dispersive Shifts} \label{sec:Dispersive_Shifts}
In the main text we described how we derived the dispersive shifts and gave some expressions relating the bare and true dispersive shifts. The expressions for these shifts get larger the more qubits are added to the system. We have included a link to a repository with a Mathematica notebook where these expressions can be generated and explored \cite{Bakergithub2022}.
\section{Parameters}
Here we give some example parameters which would allow for the tessellation shown in Figure \ref{fig:lattice_and_diagram}. Since there are many other parameters associated with the lattice namely the coupler frequencies which will affect the couplings between the qubits, we only give the qubit frequencies. These frequencies have been shown to give dispersive shifts as described in the main text with all other (higher order and unwanted) shifts suppressed to below $0.5 $MHz.

\begin{figure}[ht]
    \begin{center}
        \begin{tabular}{||c c c||} 
         \hline
         Qubit & Transition Frequency $[GHz]$ & Anharmoncity $[GHz]$ \\ [0.5ex] 
         \hline\hline
         $Q_1$ &  5.28 & -0.13\\
         $Q_2$ &  5.35 & -0.15\\
         $Q_3$ &  5.43 & -0.27\\
         $Q_4$ &  5.5 & -0.26\\
         $Q_5$ &  5.24 & -0.26\\
         $Q_6$ &  5.4 & -0.2\\
         $Q_7$ &  5.31 & -0.22\\
         $A_1$ & 5.02 & -0.17\\
         $A_2$ & 4.96 & -0.21\\
         $A_3$ & 5.06 & -0.27\\
         $A_4$ & 4.93 & -0.2\\ 
         \hline
        \end{tabular}
    \label{tab:lattice_parameters1}
    \caption{Table outlining the parameters that provide equal dispersive shifts within each unit cell as described in Fig. \ref{fig:SW_progression}. The rest of the dispersive shifts are higher suppressed. Here we have labelled the ancilla qubits as $A_i$ and the qubits as $Q_i$.}
    \end{center}
\end{figure}

\section{Dispersive Shifts} \label{sec:Appendix_DispersiveShifts}
We show the all of the dispersive shifts that were calculated in our simulations these shifts were tuned within reason so that the unwanted shifts were small enough and the data qubit - ancilla shifts were equal and dominant. We give each shift to three decimal places.

\begin{figure}[ht]
    \begin{center}
        \begin{tabular}{||c c||} 
         \hline
         Dispersive Shift & Strength [MHz]  \\ [0.5ex] 
         \hline\hline
         $\chi_{12}$ &  -5.005  \\
         $\chi_{13}$ &  -5.079  \\
         $\chi_{14}$ &  -5.050  \\
         $\chi_{23}$ &  0.030  \\
         $\chi_{24}$ &  -0.212  \\
         $\chi_{34}$ &  0.079  \\
         $\chi_{123}$ &  0.246  \\
         $\chi_{124}$ &  0.359  \\
         $\chi_{134}$ &  0.072  \\
         $\chi_{234}$ &  -0.024  \\
         $\chi_{1234}$ &  0.002  \\
         \hline
        \end{tabular}
    \label{tab:lattice_parameters2}
    \caption{Complete list of dispersive shifts in our simulations.}
    \end{center}
\end{figure}
\newpage
\bibliography{references}
\end{document}